\documentclass[aps,prx,twocolumn,amsmath,amssymb,superscriptaddress,floatfix,nofootinbib]{revtex4-2}

\usepackage{graphicx}
\usepackage{amsfonts}
\usepackage{booktabs}
\usepackage{xcolor}
\usepackage{verbatim}

\newcommand{\rev}[1]{\textcolor{black}{#1}}
\definecolor{revtwo}{HTML}{B8541D}
\newcommand{\revn}[1]{\textcolor{black}{#1}}

\begin{document}

\title{Autoregressive Neural Network Extrapolation of Quantum Spin Dynamics Across Time and Space}

\author{Hubert Pugzlys}
\affiliation{Department of Physics, University of Florida, 2001 Museum Road, Gainesville, FL 32611, United States}
\author{Shreyas Varude}
\affiliation{Department of Physics, University of Florida, 2001 Museum Road, Gainesville, FL 32611, United States}
\author{Sam Dillon}
\affiliation{Department of Physics, University of Florida, 2001 Museum Road, Gainesville, FL 32611, United States}
\author{Huy Tran}
\affiliation{Department of Physics, University of Florida, 2001 Museum Road, Gainesville, FL 32611, United States}
\author{Ta Tang}
\affiliation{Department of Applied Physics, Stanford University, 348 Via Pueblo Mall, Stanford, CA 94305, United States}
\affiliation{Stanford Institute for Materials and Energy Sciences, SLAC National Accelerator Laboratory, 2575 Sand Hill Road, Menlo Park, CA 94025, United States}
\author{Brian Moritz}
\affiliation{Stanford Institute for Materials and Energy Sciences, SLAC National Accelerator Laboratory, 2575 Sand Hill Road, Menlo Park, CA 94025, United States}
\author{Zhe Jiang}
\affiliation{Department of Computer and Information Science and Engineering, University of Florida, 432 Newell Dr, Gainesville, FL 32611, United States}
\author{Chunjing Jia}
\email{cjia1@ufl.edu}
\affiliation{Department of Physics, University of Florida, 2001 Museum Road, Gainesville, FL 32611, United States}

\begin{abstract}
Understanding the long-time dynamics of strongly correlated quantum matter remains a major challenge: as quantum entanglement grows during time evolution, conventional numerical methods become increasingly costly, limiting access to the long-range correlations and low-energy excitations that govern gapless systems. Here we show that these dynamical correlations can be extended beyond the accessible numerical window without explicitly evolving the underlying many-body wavefunction. 
We introduce an autoregressive machine-learning framework that enables the extrapolation of dynamical spin correlations in both time and space beyond the reach of conventional numerical methods. Trained on time-dependent density matrix renormalization group simulations of the gapless XXZ model, our approach is benchmarked against exact solutions available for this analytically solvable system. Combined with physics-informed spatial extension, a multi-layer perceptron model using ReLU activation functions has been shown to be superior to convolutional neural networks and linear regressions for longer time extrapolation. A study of error accumulation further demonstrates that our autoregressive neural network extrapolations are highly robust to perturbations, suggesting stable and reliable predictions. This work \revn{offers a complementary route to studying} the dynamics of gapless quantum many-body systems, in which machine learning extends and complements the capabilities of state-of-the-art numerical approaches.
\end{abstract}

\maketitle

\section{Introduction}
\label{Sec:Intro}

Machine learning (ML) has rapidly emerged as a transformative tool in condensed matter and materials physics, enabling advances in identifying phase transitions and recognizing phases of matter \cite{van2017learning,carrasquilla2017machine,zhu2024active,broecker2017quantum,PhysRevB.99.121104,PhysRevB.94.195105,PhysRevE.96.022140,doi:10.1126/science.abk3333}, predicting new materials \cite{PhysRevLett.120.145301,doi:10.1021/acs.chemmater.9b01294,choudhary2022recent,schmidt2019recent,wei2019machine,liu2024self}, developing scalable methods for quantum state tomography and simulation \cite{torlai2018neural,quek2021adaptive,schmale2022efficient,PhysRevA.106.012409,PhysRevA.102.042604}, and many others \cite{RevModPhys.91.045002}. In particular, neural-network quantum states (NQS) have been introduced as powerful variational representations of many-body wavefunctions \cite{carleo2017,Sharir2020}, extending the reach of classical methods to regimes beyond those accessible to density matrix renormalization group (DMRG) approaches. Despite this progress, simulating the dynamics of strongly correlated quantum systems remains a central challenge. Growth of entanglement entropy during time evolution, predicted by the Lieb–Robinson bound \cite{lieb1972finite}, quickly drives up the computational cost, while numerical errors such as Trotterization accumulate at long times \cite{PhysRevLett.93.076401,PhysRevB.77.134437,tebd,FOREST1990105,OMELYAN2002188}. These barriers have so far limited our ability to probe dynamical responses in large systems and over extended time scales.

\begin{figure}[t]
    \centering
    \includegraphics[width=\columnwidth]{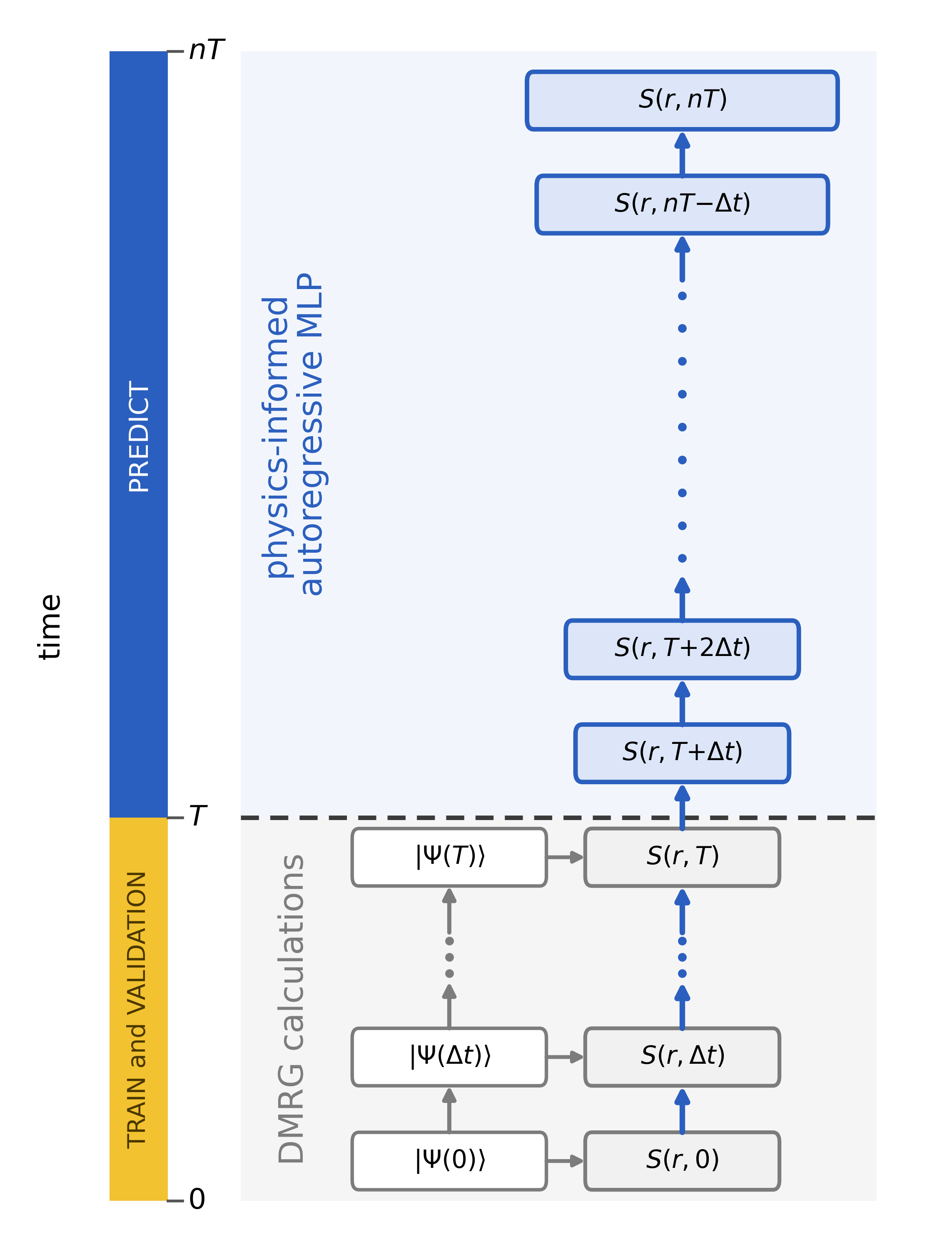}
\caption{\rev{Schematic of the proposed framework, with time running
    upward. The bar at left marks the training and validation window
    ($t\in[0,T]$, gold) and the prediction window ($t\in(T,nT]$, blue).
    Below the dashed line at time $T=20\,J^{-1}$, tDMRG evolves
    $|\Psi(t)\rangle=e^{-iHt}S^z_0|\text{G.S.}\rangle$ with
    $\hat{U}(\delta t)$, and
    $S(r,t)=e^{iE_0t}\langle\text{G.S.}|S^z_r|\Psi(t)\rangle$ is evaluated
    from each wavefunction; all tDMRG-calculated quantities and processes are
    gray. The tDMRG data within $t\in[0,T]$ is split $80/20$ at random into
    training and validation sets for the MLP. Above time $T$, the physics-informed autoregressive MLP advances
    $S(r,t)$ directly (blue), one time step at a time, to $nT=60\,J^{-1}$.
    The blue arrows in both regions represent the same learned mapping: in the training/validation region, it applies to the tDMRG ground-truth; above $T$ it operates autoregressively. }}
    \label{Fig:Schematic}
\end{figure}

Here we propose a novel strategy that sidesteps the explicit representation of many-body wavefunctions. Instead of modeling quantum states directly, we apply ML to learn directly from the dynamics, through correlation functions $S(r,t)$, which often exhibit reduced complexity in the underlying temporal or spatial patterns. \rev{Figure~\ref{Fig:Schematic} illustrates this idea: the true many-body wavefunction evolves in time through numerical simulations, but instead of tracking the wavefunction we learn the time evolution of the measurable correlation function $S(r,t)$ itself.} We formulate the task as an autoregressive prediction problem, where a series of past-time and short-distance correlations is fed into the network to predict values at the next time step and longer distances. While \emph{linear} neural networks have been recently applied to extend the time and spatial dynamics of gapped systems~\cite{tang2025improvingspectralresolutionrealtime}, our newly developed approach in this manuscript uses a \emph{nonlinear} neural network as a transfer function or autoregressive operator \cite{box2015time} to handle the long-range entanglement present in the spatial correlations of gapless systems. This framework is trained on time-dependent DMRG (tDMRG) data for the gapless XXZ model, benchmarked against Bethe ansatz results, and validated through perturbations analysis, demonstrating both accuracy and robustness. While we focus on one-dimensional spin chains here, the method is versatile and promising for higher-dimensional and more exotic systems such as quantum spin liquids.



\section{Quantum spin model Hamiltonian and numerical simulations}
\label{Sec:Results}

We focus on the one-dimensional spin-1/2 
XXZ model, a paradigmatic system of strongly correlated quantum matter that is exactly solvable and extensively studied \cite{bethe1931,giamarchi2003quantum,nagaosa1999quantum,gogolin2004bosonization,slavnov2022algebraic,korepin1997quantum,sachdev1999quantum,fradkin2013field}, and can describe the spin dynamics of various compounds \cite{PhysRevLett.70.4003,PhysRevB.52.13368,scheie2022quantum,PhysRevLett.127.037201,WitnessQuantumCorrelation}. The elementary excitations of the spin-1/2 
XXZ chain are spinons, which are fractionalized spin excitations \cite{FADDEEV1981375}. The study of fractionalized excitations is one of the most intriguing and active topics in condensed matter physics, being not only important for fundamental understanding but also holding potential applications in topological quantum computation.

The Hamiltonian for the one-dimensional XXZ model is given by
\begin{equation}
H=J\sum_{i}\left(S^x_iS^x_{i+1}+S^y_iS^y_{i+1}+\Delta S^z_iS^z_{i+1}\right),
\end{equation} 
where $S^x_i$, $S^y_i$, and $S^z_i$ are spin operators at site $i$, $J$ is the nearest-neighbor spin-exchange interaction and $\Delta$ is an anisotropy parameter. For $\left|\Delta\right|>1$, the system is gapped and the ground state exhibits an (anti)ferromagnetic Ising order for $J\Delta<0$ ($J\Delta>0$). The non-commutative nature of the quantum spin operators introduces strong quantum fluctuations in one dimension. Hence, for $\left|\Delta\right|\leq 1$, the system is gapless with vanishing ground state magnetization. In the rest of the manuscript, we focus on the gapless regime $|\Delta|\leq 1$.

Dynamical correlation functions are an important quantity in condensed matter physics that characterize how fluctuations of a physical quantity at one space-time point are correlated with fluctuations at another. Here, a dynamical correlation of the operator $S^z$ is defined as 
\begin{equation}
S(r,t) = \langle\text{G.S.}|e^{iHt}S^z_re^{-iHt}S^z_0|\text{G.S.}\rangle,
\end{equation}

\noindent where the ground state $|\text{G.S.}\rangle$ is the time-independent ground state wavefunction, and it is perturbed by the operator $S^z_0$ at the center site for a finite chain in this expression. The dynamical structure factor provides direct access to the elementary excitation spectrum of the system. It is defined as the Fourier transform of the correlation function 
\begin{align}
S(q,\omega)&=\int_{-\infty}^{\infty} dt\sum_{r}e^{i\omega t-iqr}S(r,t)\nonumber\\
&=2\pi\sum_m\left|\langle m|S_q^z|\text{G.S.}\rangle\right|^2 \delta\left(\omega-E_m + E_0\right).
\label{Eq:DSF_Definition}
\end{align}
where $|m\rangle$ and $E_m$ are the excited states and the corresponding energies ($E_0$ for the ground state); $S^z_q$ is the Fourier transform of the real space spin operator $S^z_r$. 

In our study, the dynamical correlation function $S(r,t)$ for the XXZ model is simulated by tDMRG for a 200-site spin chain and times up to 60$\,J^{-1}$ using the ITensor package \cite{ITensor}. Details of the tDMRG simulation are described in Appendix.~\ref{Sec:Methods}. The tDMRG calculated data are shown within the the dashed boxes of Fig.~\ref{Fig:DataExtrapolation}(a-b), spanning over the sites $r\in[-100,100]$ and time $t\in [0,20\,J^{-1}]$, used for training and validation. The time step of the tDMRG simulation is $\delta t=0.05\,J^{-1}$. A clear light-cone structure is observed, reflecting causal propagation of excitations at a certain velocity. The imaginary part of the correlation, $\text{Im}\left[S(r,t)\right]=\langle\text{G.S.}|\left[S^z_r(t),S^z_0(0)\right]|\text{G.S.}\rangle/2i$, decays exponentially outside the light-cone. Meanwhile, the real part, $\text{Re}\left[S(r,t)\right]=\langle\text{G.S.}|\left\{S^z_r(t),S^z_0(0)\right\}|\text{G.S.}\rangle/2$, contains important correlations outside the light-cone, with decaying and oscillating behavior as a result of low energy spinon-antispinon excitations near momentum $\pi$. This tDMRG calculated $S(r,t)$ will serve as the training set for our ML approach as described in the next section. 

\section{Neural Network Autoregressive framework for spatial and temporal extension}

\subsubsection{Two-stage Autoregressive Framework for Gapless Systems}
An autoregressive architecture with linear regression was developed and applied to gapped systems~\cite{tang2025improvingspectralresolutionrealtime}. In contrast, here we use a neural network model that tackles the much more challenging case of gapless systems, where long-range correlation leads to power-law decay outside the light-cone rather than the exponential decay seen in gapped systems. Capturing this behavior is crucial in quantum critical systems, underscoring the broader significance of our approach. \rev{The overall strategy is summarized in Fig.~\ref{Fig:Schematic}. Instead of time-evolving the many-body wavefunction, our ML framework learns to evolve the observable directly: a neural network acts as an effective, data-driven time-evolution operator that maps $S(r,t)$ from earlier time steps to the next (blue arrows in Fig.~\ref{Fig:Schematic}), without reconstructing the underlying many-body wavefunction.}

\begin{figure*}[t]
    \centering \includegraphics[width=0.85\linewidth]{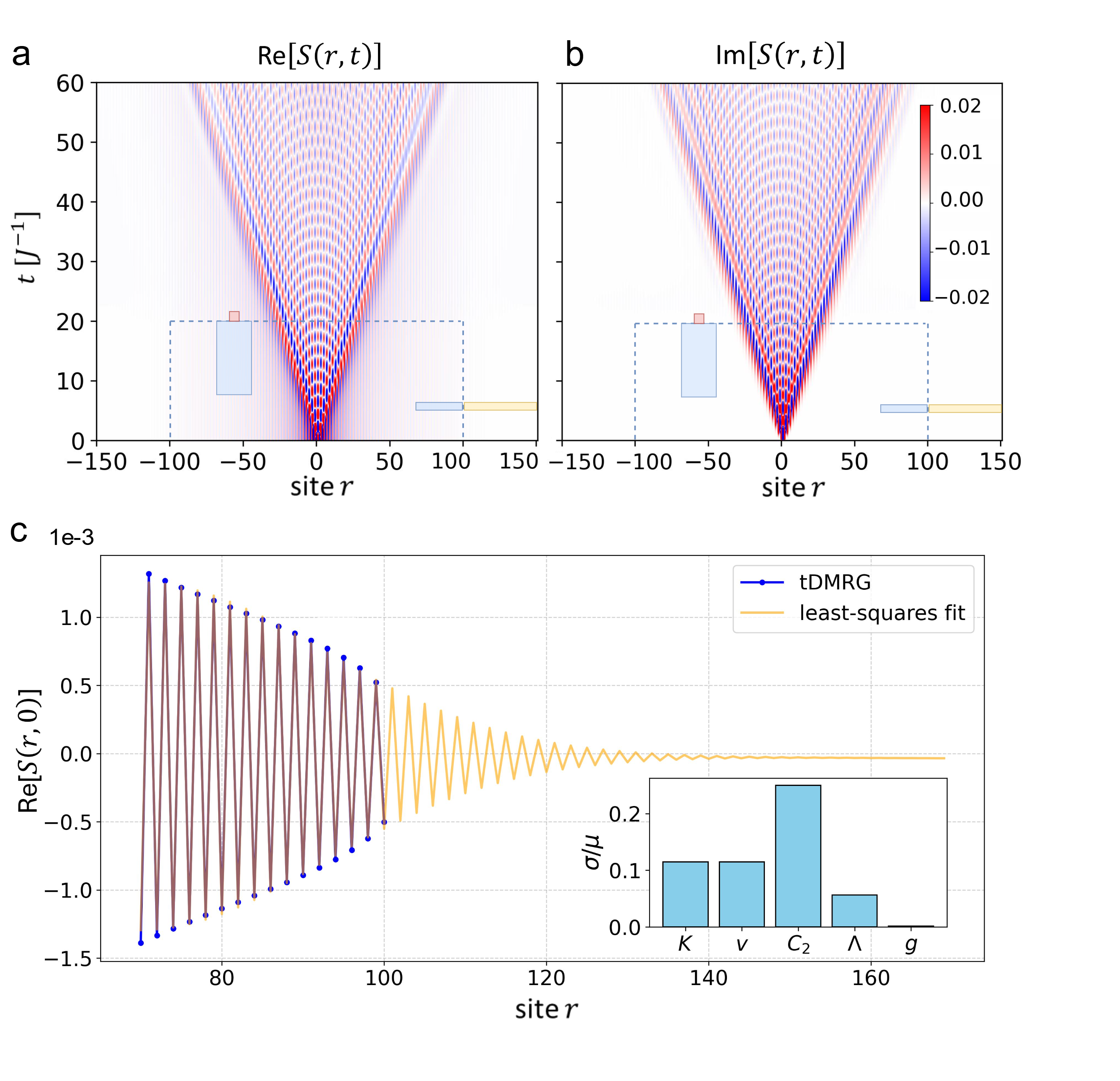}
    \caption{(a-b) The real and imaginary part of the dynamical spin correlation function $S(r,t)$ for XXZ model with $\Delta = 0.75$. 
    The tDMRG data is shown within the dashed region, while the rest of data beyond the dashed region were produced using the ML autoregressive framework based on a MLP with ReLU activation functions. The blue rectangles denote the input data for the MLP, which outputs the real and imaginary parts of a single point in $S(r,t)$ as shown with the red box. The blue and yellow rows denotes the input and output of the spatial extension. (c) tDMRG data and spatial extension based on the least-square fit to the power-law decay of Luttinger liquid theory [Eq.~(\ref{Eq:LL_Correlation_RealSpace})] at $t=0$. Finally, the inset of (c) also contains the normalized standard deviations of the parameters used in the model fitting across time $t\in[0,20\,J^{-1}]$. These parameters correspond to the parameters in Eq.~(\ref{Eq:LL_Correlation_RealSpace}).}
\label{Fig:DataExtrapolation}
\end{figure*}

From the schematic in Fig.~\ref{Fig:Schematic}, Fig.~\ref{Fig:DataExtrapolation} illustrates our new autoregressive ML framework to extrapolate spin dynamics in both time and space. The tDMRG algorithm provides spin dynamics within the dashed box ($r\in[-100,100]$, $t\in[0,20\,J^{-1}]$), and then the framework iteratively extends the spin dynamics over incremental time steps through neural network autoregressive predictions. 
Specifically, a neural network model takes spin dynamics from earlier time steps as inputs ({\it i.e.} the blue boxes in Fig.~\ref{Fig:DataExtrapolation}(a-b)) and predicts the spin dynamics at the central pixel location in the next time step ({\it i.e.} the red box on top of the blue box) as the output. 
Specifically, to predict a single complex $S(r,t)$ value, the model takes as input a finite local window of data spanning $[r-l, r+l]$ in space and $\left[t-(h+1)\,\delta t,t-\delta t\right]$ in time, where the temporal resolution is $\delta t = 0.05\,J^{-1}$; $l$ and $h$ are hyperparameters representing the window size along the spatial and temporal dimensions, respectively. The model outputs two numbers corresponding to $\text{Re}\left[S(r,t)\right]$ and $\text{Im}\left[S(r,t)\right]$. 
By sliding the window in space across the most recent time step of the tDMRG data, we extrapolate a new time step for all sites. 

To mitigate potential boundary effects during temporal extensions—arising from the extrapolation of the light cone to the edge sites of the dashed region ({\it i.e.} when the red pixel falls right above the upper left corner of the dashed region)—we first apply a spatial model to extend (pad) the input data in both the left end and the right end of the spin chain. 
For the spatial extension outside of the light cone, we utilized a physics-informed power-law least-squares regression model, as predicted by Luttinger liquid theory \cite{giamarchi2003quantum,sachdev1999quantum}. The analytical expression for the spatial dependence of the correlation function of a Luttinger liquid is summarized in Appendix~\ref{Sec:LuttingerLiquid}. As the power-law applies to regions far from the light cone, we use the last 30 sites from the boundary to train the regression model. Figure~\ref{Fig:DataExtrapolation}(c) shows the spatial extrapolation at the initial time step $t=0$ in the region away from the light cone. There is a good agreement between the simulated (blue) and the fitted (yellow) data. The regression model, Eq.~(\ref{Eq:LL_Correlation_RealSpace}), can be fit at different times, giving us a set of fitted regression parameters. The inset of Fig.~\ref{Fig:DataExtrapolation}(c) shows that the standard variation $\sigma$ of the fitted parameters across time  $\in[0,20\,J^{-1}]$ remains low compared to their average values $\mu$. Indeed, the universal parameters, such as Luttinger parameter $K$ and Fermi velocity $v$, have a low relative standard variation. Meanwhile, the relative variance of the non-universal parameter $C_2$ is comparatively large. Given the low standard variation in regression parameters across models trained at different times, a single regression model trained at $t=0$ is employed for the spatial extension for all time steps.

Our autoregressive framework runs as follows. Given the precomputed spin dynamics within the dashed region, it first conducts spatial extension to pad the spin dynamics to the left and right of the dashed region ({\it i.e.} $r\in[-150,-100]\cup[100,150]$ for $t\in[0,20\,J^{-1}]$). Then, for each consecutive time step, the neural network model is employed to predict the $S(r,t)$ values at all spatial locations (except near the boundary) by scanning the blue window horizontally. The $S(r,t)$ values can be padded by the spatial extension model. This autoregressive approach is then repeated, using the previously predicted time steps as inputs to the NN model, to extrapolate for the next time step.

The specific neural network model architecture we chose is a multi-layer perceptron (MLP) with rectified linear unit (ReLU) as the non-linear activation. The spin dynamic values within the window are flattened into a vector with $2(2l+1)h$ values as MLP inputs.  We also tried other model architectures, such as the convolutional neural network (CNN) and a linear regression model, as baselines for comparison.

\subsubsection{Model Implementation}

{\bf Ground truth spin dynamics computation.} In designing the model, we identified a few key factors to effectively learn the underlying dynamics. First, enough time steps are required for accurate learning of the spin dynamics. In the current case, training the model from a dataset simulated to less than $t=20\,J^{-1}$ reduces its ability to preserve the light cone at longer time steps. This limitation arises because the cone structure constitutes a smaller proportion of the input data at earlier time steps. Meanwhile, the information of the spin dynamics (including those at short distances) is mostly encoded within the light cone itself. Therefore, enough time steps are needed. 
As such, this constrained the minimum number of time steps needed for long-time accurate temporal extension. Considering the temporal resolution is $\delta t=0.05\,J^{-1}$, the ground truth spin dynamics are precomputed within a $200$ by $400$ dash box for training/validation, as shown in Figs.~2(a) and (b).

{\bf Neural Network Model Implementation.}
The neural network used to extend the data in time is a MLP. The model consists of five fully connected layers: an input layer, three hidden layers with 256, 128, and 64 neurons, respectively, and a two-dimensional output layer representing the real and imaginary components of a single complex value. The input layer has dimension $2(2\ell + 1)h$, where $2\ell+1$ and $h$ are hyperparameters specifying the width and height of each input data window. The ReLU activation function is applied to all hidden layers. Training was performed using the Adam optimizer with an initial learning rate of $10^{-4}$. The ``speed of light" for the light-cone aids the design of the hyper-parameters $(\ell,h)$ of the box of the input data. The ideal hyper-parameters found through grid search are $h=5.5\,J^{-1}$ and $\ell=3$ for $\Delta = 0.25$ and $h=5.5\,J^{-1}$ and $\ell=4$ for $\Delta = 0.75$. The choice of the box width/height ratio is also consistent with the shape of the light cone for each $\Delta$. For $\Delta = 0.75$, the neural network comprises approximately 295,000 trainable parameters.

{\bf Training, testing and validation.} For neural network training and validation, the ground truth tDMRG data is partitioned by sliding overlapping windows across the symmetrized correlation function, which occupies $401$ time rows and $199$ sites. Sliding the $9\times55$ input box gives $346\times191=66\,086$ distinct (box, target) pairs, where each target is the single complex value at row $i+55$ and column $k+4$ for a box whose bottom-left corner sits at row $i$ and column $k$; the outermost four sites on each side and the first $55$ time rows cannot serve as targets. Every pair has its target inside the tDMRG window ($t\le20J^{-1}$), and all of them are used: the full set is divided $80/20$ at random into $52\,869$ training and $13\,217$ validation windows, discarding none.

The validation set is utilized for the learning-rate schedule and early-stopping criterion during training. All quantitative results reported below are evaluated on $t\in(20,60]J^{-1}$ against a tDMRG calculation carried to $t=60J^{-1}$ for testing; that interval lies beyond the input window and is touched by neither training nor validation data.

\vspace{4mm}
\noindent
\section{Results}

\begin{figure}[t]
    \centering
    \includegraphics[width=\linewidth]{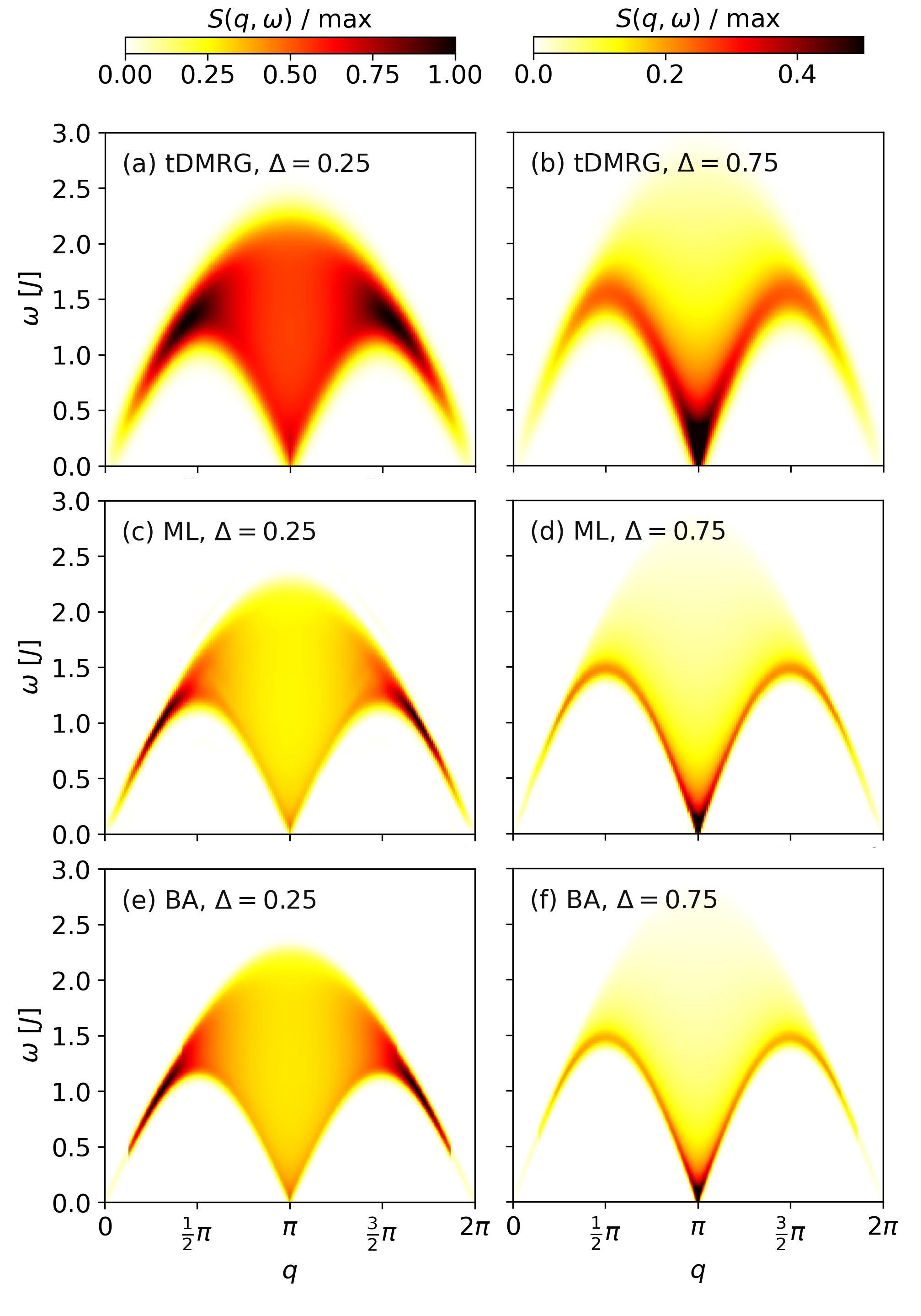}
    \caption{\rev{Dynamical structure factor $S(q,\omega)$ of the spin-$1/2$ XXZ chain for $\Delta=0.25$ (left column) and $\Delta=0.75$ (right column). (a,b) tDMRG data alone, Fourier transformed over $t\in[0,20\,J^{-1}]$ and 200 sites. (c,d) after our physics-informed autoregressive extrapolation to $t\in[0,60\,J^{-1}]$ and 300 sites. (e,f) Bethe ansatz (BA) reference obtained with ABACUS \cite{ABACUS2009}, taken as the ground truth for the one-dimensional XXZ chain. All panels are normalized to their own maximum and share a single color scale.}}
    \label{Fig:Improved_DSF}
\end{figure}

\subsubsection{Improved Resolution of the Dynamical Structure Factor and Benchmarking against the Bethe Ansatz}

As shown in Fig.~\ref{Fig:DataExtrapolation}(a,b), using our autoregressive framework based on MLP with ReLU activation functions, we extended $S(r,t)$ from $t\in [0,20\,J^{-1}]$ and 200 sites to $t\in[0,60\,J^{-1}]$ and 300 sites. The ML extrapolated results show smooth transitions from the tDMRG data within the dashed boxes for both real and imaginary parts of the correlation function, capturing all features within and outside of the light-cones. \rev{Fourier transforming the tDMRG data alone, from time to energy and from real space to momentum space, results in a blurred spectrum for both anisotropies [Figs.~\ref{Fig:Improved_DSF}(a,b)]. In contrast, the Fourier transform of the ML extrapolated correlation function yields much sharper spectra [Figs.~\ref{Fig:Improved_DSF}(c,d)], in which the momentum $q$ carrying the highest intensity shifts closer to the $\Gamma$ and $2\pi$ points for $\Delta=0.25$ and sharpens to the lower edge of the two-spinon continuum and the $\pi$ point for $\Delta=0.75$.}

\begin{figure*}[t]
        \centering
{\includegraphics[width=0.75\linewidth]{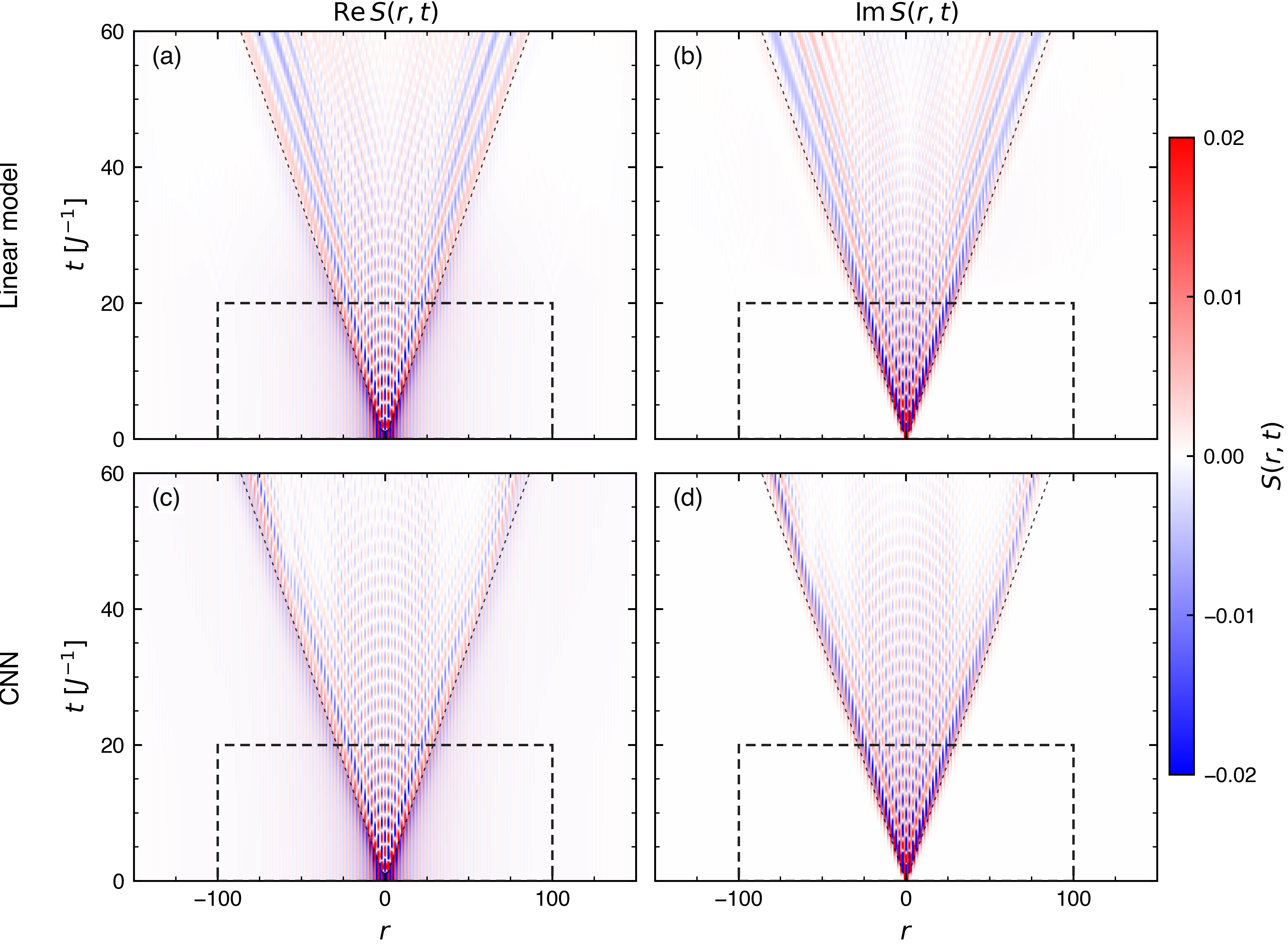}}
    \caption{%
Autoregressive extrapolation of the dynamical spin correlation function
$S(r,t)$ for the XXZ chain at $\Delta=0.75$: (a,b) the linear model and
(c,d) the CNN, showing the real part (a,c) and the imaginary part (b,d).  The dashed rectangle marks the tDMRG input region; everything outside it is produced by the
autoregressive extrapolation and Luttinger-liquid power-law extension. The thin dotted lines show the
light cone from the ground truth.}
    \label{Fig:Best4}
\end{figure*}

\rev{To establish that this sharpening reflects the true dynamics rather than an artifact of the extrapolation, we benchmark against the Bethe ansatz, which provides the exact solution of the one-dimensional XXZ chain. We solved the Bethe ansatz numerically using ABACUS \cite{ABACUS2009}; the results are shown in Figs.~\ref{Fig:Improved_DSF}(e,f). For both $\Delta=0.25$ and $\Delta=0.75$, the ML extrapolated spectra agree with the Bethe ansatz reference, whereas the tDMRG-only spectra do not resolve the refined low-energy features that both of the former exhibit. This agreement demonstrates that the ML framework extends $S(r,t)$ accurately in both space and time, and that the gain in resolution is physical.} 

\rev{In particular, we clearly observe a transfer of spectral weight toward the low-energy excitations near momentum $q\sim\pi$ upon increasing the anisotropy parameter $\Delta$ [Figs.~\ref{Fig:Improved_DSF}(c,d)]. A larger anisotropy parameter means stronger interaction and thus stronger Umklapp scattering strength at zero magnetization. The observed spectral weight transfer, together with the interpretation of the anisotropy parameter as an interaction strength, indicates the onset of anti-ferromagnetic order when $\Delta>1$. Our ML method captures this spectral weight transfer in the dynamical structure factor, consistent with expectations from the underlying physics and the Bethe ansatz reference in Figs.~\ref{Fig:Improved_DSF}(e,f).}


\subsubsection{Quantitative Model Evaluation and Extrapolative Model Performance}

To evaluate the predictive performance of the MLP, we compared it with several alternative NN architectures, including a CNN and a linear regression model (a dense layer without non-linear activation). 
We focus on the extrapolation of the same tDMRG data for the spin-$1/2$
XXZ chain for $\Delta = 0.75$. Each model is trained on a fixed input box
of $\ell = 4$ ($2\ell+1 = 9$ sites) $\times\, h = 55$ time steps taken from the raw $\delta t = 0.05\,J^{-1}$ grid;
the input is symmetrized, $S(r,t)\!\to\![S(r,t)+S(-r,t)]/2$, and the chain
is padded on both sides by the spatial extension model using Luttinger-liquid power-law decay as discussed in Section III.1. The architectures of the linear model and CNN, both trained with a learning-rate schedule and early stopping, are described as follows.

\textbf{Linear model.} A single dense layer ($1{,}982$ parameters, no activation) trained with initial learning rate $\eta_0 = 10^{-4}$ and a \texttt{ReduceLROnPlateau} schedule (factor $0.5$, patience $8$, $\eta_{\min} = 10^{-7}$), with early stopping (patience $30$, restore-best weights) for up to $400$ epochs.

\textbf{CNN.} A convolutional network ($196{,}738$ parameters) treating the input box as a two-channel ($\mathrm{Re},\mathrm{Im}$) image: two $3\times3$ convolutions ($32$ channels) $\to$ $2\times2$ max-pool with stride 1 $\to$ two $3\times3$ convolutions ($64$ channels) $\to$ adaptive average pool to $4\times4$ $\to$ a $128$-unit dense head. Trained with initial learning rate $\eta_0 = 10^{-4}$ under the same \texttt{ReduceLROnPlateau} schedule and early-stopping settings for up to $400$ epochs; the reported weights are the restore-best (lowest validation loss) weights.

For all architectures, training uses the random $80/20$ validation split described in Section~III and the Adam optimizer with mini-batch size $32$.

\begin{table*}[t]
\caption{Extrapolation metrics on $t\in(20,60]J^{-1}$ for the three architectures, all trained with the random $80/20$ validation split, a learning-rate schedule and early stopping, three random seeds each.}
\label{tab:best_config}
\begin{ruledtabular}
\begin{tabular}{lrccccc}
Model & Parameters & MSE & MAE & Pearson & Asymmetry & $L^2$ vs BA \\
\hline
MLP & 294\,978 & {\boldmath$(6.30\pm3.05)\times10^{-7}$} & {\boldmath$(4.26\pm0.70)\times10^{-4}$} & {\boldmath$0.949\pm0.030$} & {\boldmath$0.053\pm0.032$} & {\boldmath$0.032\pm0.001$} \\
Linear & 1\,982 & $(1.98\pm0.04)\times10^{-6}$ & $(1.01\pm0.02)\times10^{-3}$ & $0.820\pm0.004$ & $0.054\pm0.017$ & $0.056\pm0.001$ \\
CNN & 196\,738 & $(2.64\pm1.31)\times10^{-6}$ & $(8.31\pm1.88)\times10^{-4}$ & $0.788\pm0.088$ & $0.260\pm0.105$ & $0.040\pm0.005$ \\
\end{tabular}
\footnotetext[1]{Entries are mean $\pm$ s.d.\ over three independent random seeds, which set the weight initialization, the training/validation assignment and the mini-batch order. The same three seeds are used for every architecture. Boldface marks the best mean in each column. }
\end{ruledtabular}
\end{table*}


The predictive performance of the architecture by implementing the CNN and linear models is shown in Fig.~4. 
The linear model with learning rate schedule and early stopping has good performance for long-time extrapolation. However, while CNN has overall stable long-time extrapolative performance, it predicts non-symmetric correlations at long times, which is not physical.

\begin{figure}[t]
        \centering
        {\includegraphics[width=\linewidth]{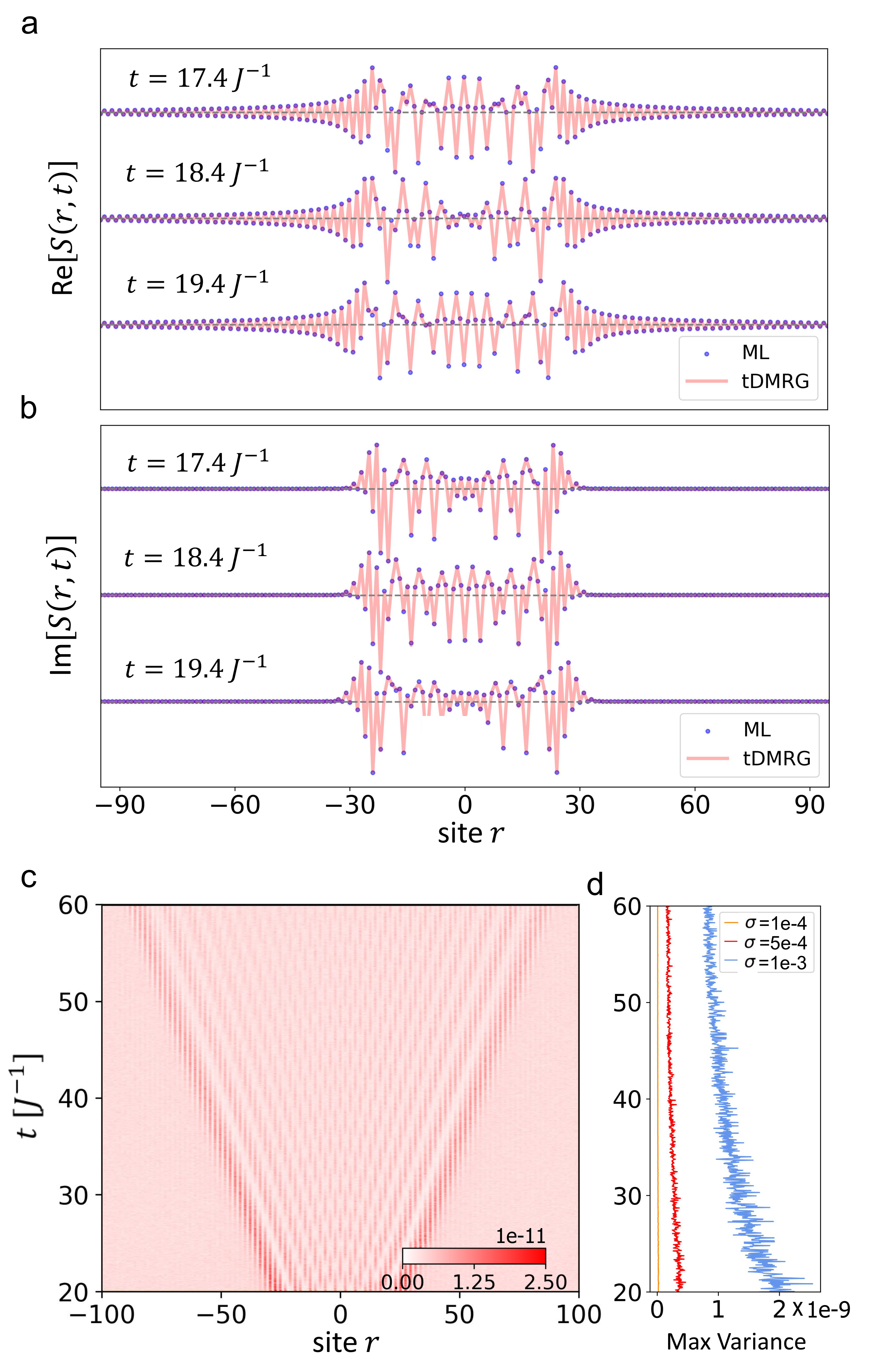}}
    \caption{Perturbative study on error accumulation effect. A set of 1000 randomly perturbed MLP models is generated for extrapolation. (a-b) The average extrapolation value of Re$\left[S(r,t)\right]$ and Im$\left[S(r,t)\right]$, overlapped with test data produced by tDMRG at finite time steps. Evidently, the ML prediction (blue dot) quantitatively matches with the test data (red line). (c) Variance of $S(r,t)$ of the perturbed models over time and space. The gaussian noise is set to $\sigma=10^{-4}$. (d) The maximum variance at each time step with varying magnitudes of gaussian noise decreases with time. 
    }
    \label{Fig:Error}
\end{figure}

To quatitatively compare the performance of MLP with linear model and CNN, we evaluated a few metrics for the predictive performance, as shown in Table I. These metrics include: (1) mean-squared error (MSE), (2) mean absolute error (MAE), and (3) Pearson correlation coefficients over the extrapolated region of
$t \in (20, 60]J^{-1}$ (real and imaginary parts, versus the tDMRG calculated ground
truth), (4) the integrated reflection asymmetry (Asym)
$\sum_{r>0,t} |S(r,t) - S(-r,t)| / \sum_{r>0,t} |S(r,t) + S(-r,t)|$ over the
extrapolated region $t \in (20, 60]J^{-1}$, and (5) the relative $L^2$ distance of the
Fourier-transformed dynamical structure factor $S(q,\omega)$ to the
Bethe-ansatz calculated $S(q,\omega)$ over $0.2 \le q/\pi \le 1.8$. The Pearson correlation coefficient is good at capturing the pattern matching, while the $L^2$ distance captures the absolute differences in spectral weights. The metrics shown in Table I indicate that CNN shows the worst asymmetry behavior, and MLP shows the better Pearson correlation coefficients together with the better $L^2$ distance, compared to linear model with learning rate schedule and early stop. Overall, MLP shows a superior extrapolative performance compared with other architectures. This indicates that the MLP with ReLU activation functions likely provides a more suitable inductive bias than the other architectures for capturing the underlying physical structure of the spin dynamics. 

\noindent
\subsubsection{Uncertainty Quantification: Perturbation Study of Error Accumulation}
\label{Sec:Robustness}

In this section, we conducted a perturbation study to simulate the impact of error accumulation in the neural network-based autoregressive extrapolation framework. We focus the uncertainty quantification on the MLP architecture, which shows the best performance. Specifically, in our fully connected feedforward architecture, the neurons in each dense layer
are rescaled by a factor of $1+\epsilon^{(n)}_{ij}$, where the noise $\epsilon^{(n)}_{ij}$ is drawn independently from a Gaussian distribution, $\mathcal{N}(0,\sigma^2)$, with zero mean and standard deviation $\sigma$. In this way, we can analyze the impact of error accumulation both in our neural network model and in our autoregressive framework for spatial and temporal extrapolation. 

    Figures~\ref{Fig:Error}(a-b) compare the mean prediction from the set of noisy models and the tDMRG simulation. We plot the real and imaginary parts of the correlation function at a few typical time slices. We see that the mean prediction of noisy models matches very well with the tDMRG data, with only minor deviations. Figures~\ref{Fig:Error}(c-d) show the variances of the prediction from the noisy models in the time interval $t\in [20,60]J^{-1}$. With a noise of $\left(\epsilon^{(n)}_{ij}\right)^2\sim\sigma^2=10 ^{-8}$, the variance in the neural network prediction is extremely small $\sim 10^{-11}$, as shown in Fig.~\ref{Fig:Error}(c). Figure~\ref{Fig:Error}(d) presents the maximum variance at each instant of time for various noise levels. Surprisingly, the maximum variance decreases with time. This fact indicates that either errors do not accumulate or they accumulate at an extremely low rate. In both cases, our result suggests that the predictions from the MLP extrapolation are less impacted by the error accumulation effect. 
    With certain noise introduced to the neural network model, we show that the MLP prediction does not deviate much from the tDMRG data. More interestingly, we do not observe significant error accumulation at the latest times of the extrapolation. 

\section{Conclusion and Discussion}
\label{Sec:conclusion}

In this work, we addressed a central challenge in the study of strongly correlated quantum matter: accessing dynamical correlations at times and length scales beyond the reach of direct numerical simulation. This challenge is particularly important in gapless systems, where correlations extend beyond the light cone with algebraic decay and continue to carry information about low-energy excitations. Rather than evolving the many-body wavefunction to increasingly long times, we showed that the dynamical correlation function itself can be extended by learning its local space-time evolution and incorporating the known long-distance physics of the system.

For the gapless spin-1/2 XXZ chain, our study has shown that this approach extends the dynamical correlations from $(t\leq20J^{-1})$ to $(t\leq60J^{-1})$ and over a larger spatial range while preserving the characteristic light-cone structure and long-range correlations. Most importantly, the extended dynamics reveal physical information that is obscured by the finite simulation window. The resulting dynamical structure factors exhibit substantially sharper low-energy features and resolve the two-spinon continuum in agreement with Bethe-ansatz calculations. The evolution of spectral weight toward low-energy excitations near $(q\sim\pi)$ with increasing anisotropy is also reproduced, demonstrating that the extrapolation preserves nontrivial interaction-dependent physics rather than simply extending a numerical pattern. The stability of the predictions under controlled perturbations provides further evidence that the long-time dynamics do not suffer from significant error accumulation. Taken together, these results suggest that the accessible correlation function contains enough structured information to extend the dynamical window without explicitly reconstructing the full many-body state.

More broadly, our results point to a complementary way of studying quantum dynamics: instead of asking how far a many-body wavefunction can be evolved, one can ask how far the physically relevant observables can be continued while respecting their known constraints and asymptotic behavior. This perspective may be particularly valuable for gapless and quantum-critical systems, where long-range correlations are both difficult to simulate and central to the physics. Extending this framework to other strongly correlated systems, higher dimensions, and regimes with richer collective excitations will be an important direction for future work.


\section*{Data Availability}

The simulation code and data is available upon reasonable request.

\section*{Acknowledgements}

We thank Xuzhe Ying for his invaluable insights and many stimulating discussions throughout this work. We also thank Shuyi Li and Adrian Feiguin for their helpful discussions. This work is supported by Center for Molecular Magnetic Quantum Materials, an Energy Frontier Research Center funded by the U.S. Department of Energy, Office of Science, Basic Energy Sciences under Award no. DE-SC0019330. The work at SLAC and Stanford by T.T. and B.M. was supported by the U.S. Department of Energy, Office of Basic Energy Sciences, Division of Materials Sciences and Engineering, under Contract No.~DE-AC02-76SF00515. H.P. also acknowledges the support from the Dr. Chris B. Schaffer Physics Undergraduate Research Fund. H.T. also acknowledges the support from Oegerle Scholar and University Scholar Program Scholarship. 

\section*{Author Contribution}

C.J., T.T., H.P. and Z.J. conceived the project. H.P., and S.D. performed the tDMRG calculations. H.P. performed the Bethe Ansatz calculations. S.V. conducted the exact diagonalization calculations. H.P. performed the machine learning calculations with input from Z.J., H.T., C.J., and H.P. made the figures with input from B.M. and T.T.. The manuscript was written by C.J., H.P. and B.M., with input and revisions from all authors.

\section*{Competing Interest}

We have no conflict of interest to disclose.

\bibliographystyle{apsrev4-2}
\bibliography{biblio}

\appendix
\section{Methods}
\label{Sec:Methods}

\subsection{Luttinger Liquid}
\label{Sec:LuttingerLiquid}

The low energy physics of spin-$1/2$ XXZ chain in the gapless regime $|\Delta|\leq 1$ is described by the Luttinger liquid theory. For the XXZ chain in the absence of external magnetic field, Luttinger liquid theory predicts that the dynamical correlation function exhibits a power law decay:
\begin{align}
    S(r,t)={}&-\frac{2K}{\pi^2}\frac{r^2+(vt)^2}{\left[r^2-(v t)^2\right]^2}\nonumber\\
    &+C_2(-1)^{r}\left[\frac{2/\Lambda^2}{r^2-(vt)^2}\right]^{K}+g+\cdots
    \label{Eq:LL_Correlation_RealSpace}
\end{align}
There are several features to notice. First, only two terms with the lowest power are listed above. The first term describes the propagation of excitations with momenta around $q\sim 0$. At large distance or long time, this term decays with a fixed power $\sim\text{max}[r^2,t^2]^{-1}$. The second term describes the excitations with momentum around $q\sim\pi$ (notice the oscillating factor $(-1)^{r}$). At large distance or long time, the correlation function decays as $\sim\text{max}[r^{2K},t^{2K}]^{-1}$. The power law decay depends on the details of the interaction strength, as dictated by the Luttinger paramter $K$. Second, there is a typical velocity $v$, which describes the speed of spreading of excitations. Lastly, there are non-universal parameters $C_2$ and short-distance cutoff $1/\Lambda$. Lastly, $g$ is a constant shift we introduce in the numerical fitting, which is not the prediction of the theory of Luttinger liquid.

The dynamical correlation function above dictates the following features regarding DSF. First, there are quasi-particle like excitation near zero momentum:
\begin{equation}
    \left.S(q=k,\omega)\right|_{|k|\ll\pi}\sim |k|\delta\left(\frac{\omega}{v}-|k|\right)
\end{equation}
Second, there is a continuum near momentum $q\sim \pi$ above frequency $\omega \geq v\left|q-\pi\right|$:
\begin{align}
    \left.S(q=\pi + k,\omega)\right|_{|k|\ll\pi}&\sim \text{Im}\left(k^2-\frac{\omega^2}{v^2}\right)^{K-1}\nonumber\\
    &\sim \left|k^2-\frac{\omega^2}{v^2}\right|^{K-1}\theta\left(\frac{\omega^2}{v^2}-k^2\right)
    \label{Eq:Sqw_pi}
\end{align}
DSF diverges as a power law upon approaching the boundary of excitations $\omega=v_F|q-\pi|$.

There are some features in DSF beyond LL theory. Namely, LL theory assumes a linear dispersion at low energy. However, there is always some curvature in the band structure, leading to features beyond the prediction of Luttinger liquid theory presented above. Ref. \cite{PhysRevLett.96.196405} predicts the spectral features away from momentum $q\sim 0$ or $\pi$. Ref.~\cite{PhysRevLett.96.257202} predicts the line width of the excitations near $q\sim 0$. 

\subsection{tDMRG}
We first used the DMRG method \cite{dmrg, ITensor} to find the ground state, $|\text{G.S.}\rangle$, of the spin-$1/2$ XXZ chain. The Hamiltonian of XXZ model is as follows
\begin{equation}
H = J \sum_{i\in\mathbb{Z} }\left(S^x_iS^x_{i+1}+S^y_iS^y_{i+1}+\Delta S^z_iS^z_{i+1}\right).
\end{equation}
With DMRG method, we obtained $|\text{G.S.}\rangle$ in the form of a matrix product state (MPS) \cite{mps}. We restrict the maximum bond dimension to be $\chi=200$ with a truncation error of $10^{-10}$. 

In order to calculate the dynamical spin correlation, $S(r,t)$, it's necessary to simulate the time-evolution of $|\Psi(t=0)\rangle=S^z_0|\text{G.S.}\rangle$. This was achieved using the time-evolving block decimation (TEBD) algorithm \cite{PhysRevLett.93.076401,PhysRevB.77.134437,tebd,FOREST1990105,OMELYAN2002188}. This method repeatedly applies the Trotterized time-evolution operator $\hat{U}(\delta t)=e^{-i\delta t \hat{H}}$ to the state $|\Psi(t=0)\rangle$ to achieve the states $|\Psi(\delta t)\rangle$, $|\Psi(2\delta t)\rangle$, and so on. In particular, we used the fourth order Suzuki-Trotter decomposition of $\hat{U}(\delta t)$:\cite{PhysRevB.77.134437,tebd,FOREST1990105,OMELYAN2002188}
\begin{equation}
    \begin{split}
        \hat{U}(\delta t)\approx \hat{U}_{\text{TEBD4}}(\delta t)+\mathcal{O}(\delta t^5)
    \end{split}
\end{equation}
The error in the fourth order Suzuki-Trotter decomposition scales with the time step as $\mathcal{O}(\delta t ^5)$. As detailed in Ref.~\cite{tebd}, the fourth order Suzuki-Trotter decomposed time evolution operator is constructed as:
\begin{widetext}
\begin{equation}
    \begin{split}
        \hat{U}_{\text{TEBD4}}(\delta t)={}&\hat{U}_{\text{TEBD2}}(\delta t_1)\hat{U}_{\text{TEBD2}}(\delta t_1)\hat{U}_{\text{TEBD2}}(\delta t_2)\hat{U}_{\text{TEBD2}}(\delta t_1)\hat{U}_{\text{TEBD2}}(\delta t_1)\\
        \hat{U}_{\text{TEBD2}}(\delta t)={}&e^{-i\frac{\delta t}{2}\hat{H}_{\text{even}}}e^{-i\delta t\hat{H}_{\text{odd}}}e^{-i\frac{\delta t}{2}\hat{H}_{\text{even}}}\\
        \delta t_1={}&\frac{1}{4-4^{\frac{1}{3}}}\delta t,\quad \delta t_2=\delta t-4\delta t_1.
    \end{split}
\end{equation}
\end{widetext}
Moreover, in constructing $\hat{U}_{\text{TEBD2}}(\delta t)$, the Hamiltonian is separated into two groups:
\begin{widetext}
\begin{equation}
    \begin{split}
        \hat{H}_{\text{even}}&=J \sum_{i \in \mathbb{Z}}\left(S^x_{2i}S^x_{2i+1}+S^y_{2i}S^y_{2i+1}+\Delta S^z_{2i}S^z_{2i+1}\right)\\
        \hat{H}_{\text{odd}}&=J \sum_{i \in \mathbb{Z}}\left(S^x_{2i-1}S^x_{2i}+S^y_{2i-1}S^y_{2i}+\Delta S^z_{2i-1}S^z_{2i}\right)
    \end{split}
\end{equation}
\end{widetext}
such that both of $\hat{H}_{\text{even,odd}}$ consist of commuting local operators.

We then calculated $S(r,t)$ as follows,
\begin{equation}
S(r,t) = \langle\text{G.S.}|e^{it\hat{H}}S^z_re^{-it\hat{H}}S^z_0|\text{G.S.}\rangle,
\end{equation}
where $0$ corresponds to the central site of the spin chain. That is to say, the site numbered $N/2$, where N is the total number of sites. The time-evolution was simulated in time steps, $\delta t=0.05\,J^{-1}$. For the time-evolution, we employed a maximum bond dimension of 1000 and a truncation error of $10^{-10}$.

To obtain the dynamical structure factor, we do the following. First, we symmetrize the data $S(r,t)$ given that a chain with even number of sites is studied:
\begin{equation}
    S_{\text{sym}}(r,t)=\left[S(r,t)+S(-r,t)\right]/2
\end{equation}
Second, we perform the spatial Fourier transformation:
\begin{equation}
    S_{\text{sym}}(q,t)=\sum_r e^{iqr}S_{\text{sym}}(r,t)
\end{equation}
Third, we append the data of $S_{\text{sym}}(q,t)$ at \emph{negative} time. Given the spectral decomposition of $S(q,t)$ in Eq.~(\ref{Eq:DSF_Definition}), one can obtain the data of $S_{\text{sym}}(q,t)$ for $t<0$ as the complex conjugation of the simulated data at $t\geq 0$:
\begin{equation}
    S_{\text{sym}}(q,-t)=\left[S_{\text{sym}}(q,t)\right]^*
\end{equation}
To this end, we obtain the data $S_{\text{sym}}(q,t)$ for $t\in\left[-T_{\text{max}},T_{\text{max}}\right]$ with maximum time cutoff $T_{\text{max}}$. Lastly, we perform the temporal Fourier transformation as follows:
\begin{equation}
    S_{\text{sym}}(q,\omega)=\int_{-T_{\text{max}}}^{T_{\text{max}}}dte^{-i\omega t}S_{\text{sym}}(q,t)W(t)
\end{equation}
where the time integration is limited by the maximum time cutoff $T_{\text{max}}$. To overcome the problem regarding the ``hard'' cutoff $T_{\text{max}}$, we introduce a Gaussian windowing function:
\begin{equation}
    W(t)=\exp\left[-\frac{t^2}{T_{\text{cutoff}}^2}\right]
\end{equation}
where a soft cutoff $T_{\text{cutoff}}<T_{\text{max}}$ is introduced. It is the soft cutoff $T_{\text{cutoff}}$ that limits the frequency resolution in DSF plots. The accurate neural network extrapolation of dynamical correlation function allows us to choose a large $T_{\text{cutoff}}$. Hence, the resolution of DSF is greatly enhanced.

\subsection{Bethe Ansatz calculation}

The ABACUS implementation of the Bethe ansatz is limited by the number of form factors retained in $S(q,\omega)$ \cite{ABACUS2009}; we use 95\% of the total form factors, on a 500-site chain, in contrast to the 200 sites produced by tDMRG and extended to 300 sites by the ML framework.

\end{document}